\documentclass[12pt]{article}
\input epsf.tex

\renewcommand{\theequation}{\thesection.\arabic{equation}}
\newcommand{\be}{\begin{equation}}   \newcommand{\ee}{\end{equation}}
\newcommand{\bear}{\begin{eqnarray}}
\newcommand{\eear}{\end{eqnarray}}
\newcommand{\ba}{\begin{array}}      \newcommand{\ea}{\end{array}}
\newcommand{\lae}{\begin{array}{c}\,\sim\vspace{-21pt}\\< \end{array}}

\newcommand{\CQ}{{\cal Q}}
\newcommand{\CU}{{\cal U}}
\newcommand{\CD}{{\cal D}}
\newcommand{\CL}{{\cal L}}
\newcommand{\CE}{{\cal E}}

\newcommand{\ov}{\overline}
\newcommand{\hg}{\hat{g}}

\newcommand{\yt}{\lambda_t}
\newcommand{\hh}{\lambda_h}

\def\vbr{$\vphantom{\sqrt{F_e^i}}$}

\begin{document}

\pagestyle{empty}
\begin{titlepage}
\def\thepage {}        

\title{\LARGE \bf
Self-Breaking of the Standard Model \\ [2mm] Gauge Symmetry  
\\ [1cm]}

\author{
\small\bf \hspace*{-.3cm}
 Nima Arkani-Hamed$^1$, Hsin-Chia Cheng$^2$, Bogdan
A.~Dobrescu$^{3}$, Lawrence J.~Hall$^{1}$ \\
\\
{\small {\it \hspace*{-.6cm}
$^1$Department of Physics, University of California,
Berkeley, CA 94720, USA\thanks{e-mail: arkani@thsrv.lbl.gov,
hcheng@theory.uchicago.edu, bdob@fnal.gov, ljhall@lbl.gov}}}\\
{\small {\it 
\hspace*{.4cm} Theory Group, Lawrence Berkeley National Laboratory, 
Berkeley, CA 94720, USA }}\\
\\
{\small {\it $^2$Enrico Fermi Institute, The University of Chicago,
Chicago, IL 60637, USA }}\\
\\
{\small {\it 
$^3$Theoretical Physics Department,
Fermilab, Batavia, IL 60510, USA}}\\
 }

\date{ }
\maketitle

\vspace*{-12cm}
\noindent

\parbox[t]{1.5in}{\hspace*{-.7cm}\small hep-ph/0006238 \\ 
\hspace*{-.7cm}June 20, 2000}
\hspace{2.8in}
\parbox[t]{2in}{\small FERMILAB-PUB-00/135-T \\
EFI-2000-22 \\ UCB-PTH-00/17 \\ LBL-46144}

\vspace*{12cm}

\baselineskip=18pt

\begin{abstract}

{If the gauge fields of the Standard Model propagate in TeV-size extra
dimensions, they rapidly become strongly coupled and can form scalar
bound states of quarks and leptons.  If the quarks and leptons of the
third generation propagate in 6 or 8 dimensions, we argue that the
most tightly bound scalar is a composite of top quarks, having the
quantum numbers of the Higgs doublet and a large coupling to the top
quark. In the case where the gauge bosons propagate in a bulk of a
certain volume, this composite Higgs doublet can successfully trigger
electroweak symmetry breaking. The mass of the top quark is correctly
predicted to within 20\%, without the need to add a fundamental Yukawa
interaction, and the Higgs boson mass is predicted to lie in the range
165 - 230 GeV.  In addition to the Higgs boson, there may be a few
other scalar composites sufficiently light to be observed at upcoming
collider experiments.  }

\end{abstract}
\vfill
\end{titlepage}

\baselineskip=18pt
\pagestyle{plain} \setcounter{page}{1}




\section{Introduction and Conclusions}
\setcounter{equation}{0}

The Standard Model (SM) has three main ingredients:  1) the $SU(3)_C
\times SU(2)_W \times U(1)_Y$ gauge group; 2) three generations of
quarks and leptons; 3) a Higgs doublet.  As opposed to the gauge group
and fermion representations  which may be viewed as natural low-energy
remnants of an unified  theory, the Higgs doublet is an ad hoc
addition required solely to break the electroweak symmetry and to
accommodate the observed fermion masses.  In this paper we show that
the existence of a Higgs doublet is a  consequence of ingredients 1)
and 2) provided the gauge bosons and  fermions propagate in
appropriate extra dimensions compactified at a scale in the TeV range.

Given that gauge theories are non-renormalizable in more than four
dimensions, there is need for a physical cutoff, $M_s$, above but not
far from the compactification scale. An obvious candidate for this
cutoff is the scale of quantum gravity, as would  occur if the
gravitational coupling becomes strong at a scale in the TeV range.
This may be achieved if the space accessible to Standard Model fields
is embedded in a large volume accessible only to the
gravitons~\cite{largedim}, or if there are warped extra
dimensions~\cite{Randall:1999ee}. An alternative possibility is that
the  theory becomes embedded in some other consistent ultraviolet
completion of higher-dimensional gauge theory without gravity,  while
the scale of quantum gravity is higher.  

Below the cutoff scale $M_s$, we are dealing with an effective field
theory which includes the $SU(3)_C \times SU(2)_W \times U(1)_Y$ gauge
group and three generations of fermions in compact dimensions.  The
basic idea is that the higher-dimensional gauge interactions become
strong at the scale $M_s$ and produce fermion--anti-fermion bound
states. It is very significant that, with plausible dynamical
assumptions,  the charges of the quarks and the  leptons under  the
Standard Model gauge group are such that the most deeply bound  state
which transforms non-trivially under the gauge group is a  Higgs
doublet. Thus, a composite Higgs doublet which acquires an electroweak
asymmetric vacuum expectation value could result as  a direct
consequence of the extra dimensions.

Previously, it has been shown that the combined effect of the
Kaluza-Klein (KK) modes of the gluons is strong enough \cite{ewsb} to
give  rise to a composite Higgs doublet made of the four-dimensional
left-handed top-bottom doublet and a five-dimensional top-quark
field~\cite{Cheng:1999bg}. More generally, the strong  dynamics
intrinsically associated with gauge interactions in extra  dimensions
is a good candidate for viable theories without a fundamental Higgs
doublet~\cite{general}.

Here we consider the more natural setup where a full  generation (the
``third'' one by definition) propagates in  extra dimensions of
TeV$^{-1}$ size, and the higher-dimensional  $SU(3)_C \times SU(2)_W
\times U(1)_Y$ interactions induce electroweak  symmetry breaking.  In
section~2 we study the possible bound states and symmetry  breaking
pattern of the higher-dimensional gauge dynamics  using the most
attractive channel (MAC) analysis.  A more detailed description of the
bound states using  the Nambu--Jona-Lasinio (NJL) approximation is presented
in section 3.  Remarkably enough, it turns out that the composite
Higgs doublet has a  Yukawa coupling of order one only to the
top quark.  The model includes potentially light composite scalars
other than the  Higgs boson, which could be within the reach of future
collider  experiments.

Despite the uncertainties due to the cutoff scale, we are able  to
obtain rather reliable predictions for the top and Higgs masses
because the renormalization group (RG) equations exhibit infrared
fixed points. The top mass is predicted with  a ${\cal O}(20\%)$
uncertainty, and is consistent with  the experimental value. The Higgs
boson mass is predicted  in the $165 - 230$ GeV range (section 4).

More generally, extra dimensions accessible to Standard Model fields
provide a natural setting for theories with composite Higgs fields.
Normally, in four dimensions, these theories suffer from the
difficulty that the SM Yukawa couplings look quite perturbative; even
for the top quark  $\lambda_t \sim 1$ rather than $\sim 4 \pi$. On the
other hand, in any theory with a composite Higgs, the Yukawa couplings
are expected to blow up at the  compositeness scale. This either
predicts too large a top quark mass if this  scale is low, or requires
us to push the compositeness scale up so high that  the usual
hierarchy problem fine-tune is needed to keep the Higgs 
light~\cite{Nambu,BHL}.
Theories with extra dimensions allow for a way out of this problem:
all  the fundamental higher-dimensional couplings, including the gauge
and Yukawa couplings, can be strong, while the effective four-dimensional 
couplings
can be perturbative due to a moderate dilution factor from the volume
of the extra  dimensions. More  precisely, strong dynamics can trigger
a composite Higgs to form  {\it in higher dimensions\/} with the
associated large couplings, but the  power-law running of couplings in
higher dimensions allow these couplings to reach perturbative infrared fixed
points without the need to push the  compositeness scale to grand unification
scale values.   The discussion of section~4 for the top and Higgs masses
holds in {\it any\/} such higher-dimensional theory,   with the
``composite'' boundary  conditions that the top Yukawa and Higgs
quartic couplings blow up at the  ultraviolet cutoff.   

In section~5 we mention various scenarios with three generations in
which some flavor non-universal effects prevent the up and charm
quarks from  forming deeply bound states at the scale $M_s$, while
also allowing the light quarks and leptons to obtain their masses.
Finally, we conclude with a comparison between our scenario and  the
supersymmetric extensions of the Standard Model in section 6.

\section{A Third Generation Model}
\setcounter{equation}{0}

Let us consider the Standard Model gauge group and one generation (the
``third'' one) of fermions in $D$ dimensions, where  four of them are
the usual Minkowski spacetime and $D-4$ spatial  dimensions are
compactified at a scale $1/R$ of a few TeV.  For even $D$, there is an
analogue of the four-dimensional  $\gamma_5$ matrix, $\Gamma_{D+1}$,
hence chiral fermions with  eigenvalues $\pm 1$ of $\Gamma_{D+1}$
exist.  Nonetheless, the higher-dimensional fermions have four or more
components. In order to obtain a four-dimensional chiral theory, the
extra dimensions must be compactified on an orbifold or with some
boundary conditions such that the zero modes of one four-dimensional
chirality are projected out.  We will concentrate mostly on the case
of chiral fermions in even  number of extra dimensions,  leaving the
more complicated discussion of vector-like fermions in $D\ge 5$ for
the Appendix.

We assign $SU(2)_W$ doublets with positive chirality, $\CQ_+$,
$\CL_+$, and $SU(2)_W$ singlets with negative chirality, $\CU_-$,
$\CD_-$, $\CE_-$. Each fermion contains both left- and right-handed
two-component spinors when reduced to four dimensions. We impose an
orbifold projection such that the  right-handed components of $\CQ_+$,
$\CL_+$, and left-handed components of $\CU_-$, $\CD_-$, $\CE_-$, are
odd under the orbifold ${\bf Z}_2$ symmetry and therefore the
corresponding zero modes are projected out.  As a result, the
zero-mode fermions are two-component four-dimensional  quarks and
leptons: ${\cal Q}_+^{(0)} \equiv (t,b)_L$, \  ${\cal U}_-^{(0)}
\equiv t_R$, \  ${\cal D}_-^{(0)} \equiv b_R$, \  ${\cal L}_+^{(0)}
\equiv (\nu_\tau, \tau)_L$,  ${\cal E}_-^{(0)} \equiv \tau_R$.

Given that the massless fermion spectrum (before electroweak symmetry
breaking) is a full generation of Standard Model fermions, the theory
is  obviously free from four-dimensional anomalies.  Nevertheless,
there may be $D$-dimensional anomalies because the theory is
chiral. There are no $SU(3)_C$ anomalies because the fermions have
vector-like strong interactions.  Similarly, the unbroken  $U(1)_{EM}$
is anomaly free, and the  gravitational anomaly cancels if we include
a singlet with negative chirality. (Its zero-mode can be identified as
a right-handed neutrino.) On the other hand, the  $SU(2)_W$ and
$U(1)_Y$ representations are chiral and there are $[SU(2)_W]^{D/2+1}$,
$[U(1)_Y]^{D/2+1}$ and mixed anomalies.  These $D$-dimensional
anomalies, however, can be canceled by the Green-Schwarz
mechanism~\cite{GS}. We will assume the presence of such a
Green-Schwarz counterterm in the effective Lagrangian so that the full
theory is non-anomalous. This term will not play any role in the
following discussion.

At the cutoff scale, $M_s$, the Standard Model gauge interactions are
non-perturbative and produce bound states.  Some of the scalar bound
states may have squared-masses significantly smaller than $M_s^2$, due
to the quadratic dependence on the cutoff  of their self-energies
\cite{NJL}.  We do not expect that the interactions which are  strong
in the ultraviolet exhibit confinement, because at large distance
($R< r < \Lambda_{\rm QCD}^{-1}$) only the zero modes of the gauge
fields  are relevant and the interactions are not strong. The
effective theory below $M_s$ involves both fermions and composite
scalars.  The squared-mass of the composite scalar decreases when the
strength of the attractive interaction that  produces the bound state
increases. For a sufficiently strong attractive interaction, the
squared-mass turns negative inducing  chiral symmetry breaking. 

In order to study the low-energy theory and the symmetry breaking
pattern, we need to identify the most attractive scalar
channels~\cite{Raby:1980my}.  In the one-gauge-boson-exchange
approximation, the binding strength of a $\ov{\psi}\chi$ channel is
proportional to  
\be 
\hg_3^2 {\bf T}_{\ov{\psi}}\cdot {\bf T}_\chi  +
\hg_2^2 {\bf T^\prime}_{\ov{\psi}}\cdot {\bf T^\prime}_\chi  +
\hg^{\prime 2} Y_\psi Y_\chi  
\ee 
where $\hg_3$, $\hg_2$ and
$\hg^{\prime}$ are the  six-dimensional $SU(3)_C \times SU(2)_W \times
U(1)_Y$ gauge couplings  at the cutoff scale, ${\bf T}$ and  ${\bf
T^\prime}$ are the $SU(3)_C\times SU(2)_W$ generators of the
corresponding fermion, and $Y$ is the hypercharge.  For computing the
relative strength of various channels it is convenient to use the
following identity: 
\be 
{\bf T}_{\ov{\psi}}\cdot {\bf T}_\chi=
\frac{1}{2}\left[C_2\left(\ov{\psi}\right)+C_2\left(\chi\right)
-C_2\left(\ov{\psi}\chi\right)\right] ~,  
\ee 
where $C_2(r)$ is the
second Casimir invariant for the representation $r$ of the gauge group.

The bound states which can be formed depend on the transformation of
the higher-dimensional fermions under charge conjugation.  Therefore,
we will consider separately the cases of $D=4k+2$ and $D=4k+4$ with $k
\ge 1$ integer. 

\subsection{Fermions in six dimensions ($D=4k+2$)}

We first study a six-dimensional [or more generally,
$(4k+2)$-dimensional] theory with chiral  fermions. Note that these
are dimensions larger than $M_s^{-1}$ accessible to the quarks and
leptons, and the discussion that  follows does not depend on the
existence of other dimensions which are either smaller than $M_s^{-1}$
or inaccessible to the  Standard Model fields.
 
In ($4k+2$) dimensions, the charge conjugation does not change the
chirality, in contrast with the $4k$-dimension cases.  Therefore,
${\CQ}^c_+$, ${\CL}^c_+$ still have positive chirality  and
${\CU}^c_-$, ${\CD}^c_-$, ${\CE}^c_-$ have negative chirality.  The
light bound states are $(4k+2)$-dimensional scalars, and their
constituents have the $\ov{\psi}_+ \chi_-$ form.

In Table~\ref{6d_scalars} we list all the attractive scalar channels
and the binding strength of the composite scalars in the MAC
approximation.  The higher-dimensional gauge couplings $\hg_i$ are
related to the  four-dimensional ones by the volume of the $D-4$
compact dimensions, $\hg_i = g_i \sqrt{V_{D-4}}$. We use the $SU(5)$
normalization for the hypercharge gauge coupling,  where $\hg^{\prime
2} = (3/5)\hg_1^2$. We denote the scalars transforming as the
left-handed doublet quark under the SM gauge group by $\tilde{q}$,
borrowing the notation from supersymmetry, and the scalars
transforming as $({\bf 3, 2}, -5/6)$  under SM gauge group by $X$.

\begin{table}[t]
\centering \renewcommand{\arraystretch}{1.5}\small
\begin{tabular}{|c|c|c|c|c|}\hline
\parbox[t]{1.9cm}{Composite \\ \hspace*{.3cm} scalar} & 
\parbox[t]{2.1cm}{$\, $ \\ constituents  \\ $\, $ } & 
\parbox[t]{3.8cm}{$SU(3) \times SU(2) \times U(1)$\\ \hspace*{.5cm}
 representation} & 
\parbox[t]{2.8cm}{$\, $ \\ binding strength  \\ $\, $ } & 
\parbox[t]{2.7cm}{relative binding\\ for $\hg_1=\hg_2=\hg_3$} \\
\hline \hline\vbr\vbr 
$H_{\cal U}$ & $\overline{\cal Q}_+ {\cal U}_-$
& $({\bf 1,2,} +1/2)$ &  $\frac{4}{3} \hg_3^2 + \frac{1}{15}\hg_1^2$ &
1 \\ \hline\vbr\vbr 
$H_{\cal D}$ & $\overline{\cal Q}_+ {\cal D}_-$ &
$({\bf 1,2,} -1/2)$ &  $\frac{4}{3} \hg_3^2 - \frac{1}{30}\hg_1^2$ &
$0.93$ \\ \hline\hline\vbr\vbr 
$\tilde{q}$ & $\ov{\cal Q}_+ {\cal D}^c_-$ & $({\bf {3},2,} +1/6)$ &  
$\frac{2}{3} \hg_3^2 +
\frac{1}{30}\hg_1^2$ & $0.5$ \\ \hline\vbr\vbr 
$X$ & $\ov{\cal Q}_+
{\cal U}^c_-$  & $({\bf {3},2,} -5/6)$ & $\frac{2}{3} \hg_3^2 -
\frac{1}{15}\hg_1^2$ & $0.43$ \\ \hline\vbr\vbr 
$H_{\cal E}$ &
$\overline{\cal L}_+ {\cal E}_-$ & $({\bf 1,2,} -1/2)$ &
$\frac{3}{10}\hg_1^2$ & $ 0.21$ \\ \hline\vbr\vbr 
$\tilde{q}^{\prime}$
& $\ov{\cal L}^c_+ {\cal U}_-$ &  $({\bf {3},2,} +1/6)$ &
$\frac{1}{5}\hg_1^2$ & $0.14$ \\ \hline\vbr\vbr
$\tilde{q}^{\prime\prime}$ &  $\overline{\cal L}_+ {\cal D}_-$ &
$({\bf 3,2,} +1/6)$ &  $\frac{1}{10}\hg_1^2$ & $ 0.07$ \\
\hline\vbr\vbr 
$X'$ & $\ov{\cal Q}^c_+ {\cal E}_-$  & $({\bf {3},2,}
-5/6)$ & $\frac{1}{10}\hg_1^2$ & $0.07$ \\ \hline
\end{tabular} 
\parbox{15.cm}{\small \caption{\small Attractive scalar channels in
six dimensions with chiral fermions\label{6d_scalars} } }
\end{table}

Although composite operators such as $\ov{\CQ}_+ \Gamma_\alpha \CQ_+$,
where $\alpha = 5,...,D$, are also scalars in four dimensions,
(reduced to $\ov{q}_L q_R \pm \ov{q}_R q_L$ in the two-component
spinor notation,) they belong to the vector channels in $D$
dimensions.  We make the usual dynamical assumption that Lorentz
invariance is not spontaneously broken by the strong gauge dynamics.
If these vector bound states do form, we assume that their masses are
close to the cutoff scale. Although the $D$-dimensional Lorentz
invariance is broken by the compactification, this breaking  occurs at
a scale significantly lower than the cutoff scale where  the
interactions become strong and the bound states are formed,  so it
should have little effect.

Above the compactification scale, the running of the four-dimensional
gauge couplings becomes power-law~\cite{dudas} due to the presence of
the KK modes. The convergence of the three SM gauge couplings is
accelerated. One typically finds that at the scale where the gauge
interactions become non-perturbative, the three gauge couplings become
comparable and are consistent with unification within theoretical
uncertainties~\cite{Cheng:1999fu}. Since the binding force is
dominated by ultraviolet interactions, the $SU(2)_W$ and $U(1)_Y$
interactions could be as important as the $SU(3)_C$ interaction. In
Table~\ref{6d_scalars} we also list the relative binding strength  for
all the attractive scalar channels by assuming $\hg_1=\hg_2=\hg_3$.
In order to avoid proton decay we do not invoke a  unified gauge
group, and simply assume that physics above $M_s$  preserves baryon
number. However, if there was a unified gauge group at $M_s$, then the
exchange of the additional gauge bosons would modify the binding
strength.

An inspection of Table~\ref{6d_scalars} shows that the  most deeply
bound states are  the six-dimensional $H_{\cal U}$ and $H_{\cal D}$
scalars,  which transform under the gauge group as the Standard Model
Higgs doublet.  Note that this is true for a wide range of couplings
$\hat{g}_i$; gauge coupling unification is not a necessity.  These
scalars have large Yukawa couplings to their constituents,
$\overline{\cal Q}_+ {\cal U}_-$ and  $\overline{\cal Q}_+ {\cal D}_-
$ respectively.  $H_{\cal U}$ is more strongly bound than $H_{\cal
D}$, so that it  naturally acquires a vacuum  expectation value (VEV),
breaking $SU(2)_W\times U(1)_Y$ down to $U(1)_{EM}$. Furthermore, if
the binding strength of $H_{\cal U}$ is not much larger than the
critical value where the squared-mass of $H_{\cal U}$ turns negative,
then the VEV of  $H_{\cal U}$ will be below the compactification
scale.  Hence, the zero mode of $H_{\cal U}$ plays the role of the  SM
Higgs doublet.

In the one-gauge boson exchange approximation, the squared-mass of
$H_{\cal D}$ is expected to stay positive, because of the difference
in the hypercharge interaction which also becomes strong,  though
significantly smaller than the compositeness scale.  The other
composite scalars, $H_{\cal E}$, $\tilde{q}$,  $\tilde{q}^{\prime}$,
$\tilde{q}''$, $X$, and $X'$ are not likely to be sufficiently
strongly bound for being relevant at low energies.  Therefore, we have
a compelling  picture, in which the electroweak symmetry is correctly
broken and  only the top quark acquires a large mass.  The low-energy
effective theory below $1/R$ is simply the Standard Model plus a
possible additional Higgs doublet (the zero mode of $H_{\cal D}$).

\subsection{Fermions in eight dimensions ($D=4k+4$)}

In eight dimensions (or more generally in $D = 4k+4$ with $k\ge 1$)
with chiral fermions, there are some different bound states because
charge conjugation flips the chirality.  Besides $H_{\cal U}$,
$H_{\cal D}$, $H_{\cal E}$, and $\tilde{q}''$, there are four more bound
states transforming like the right-handed down-type quark under the SM
gauge transformation (see Table~\ref{8d_scalars}).  Among them, the
bound state $\tilde{b}=\ov{\CQ}_+ \CQ^c_-$  is also strongly bound and
in the MAC approximation would have the same binding strength as
$H_{\cal U}$ if all three SM gauge couplings had the same
strength. The degeneracy is accidental and will not be exact.  For
example, by taking into account the effect of running couplings, the
$\ov{\CQ}_+ \CQ^c_-$ channel will be somewhat weaker than the Higgs
channel $\ov{\CQ}_+ \CU_-$ even if we assume $\hg_3=\hg_2=\hg_1$ at
the cutoff scale, because the contributions coming from scales below
$M_s$ have $\hg_2<\hg_3$. Nevertheless, the composite scalar
$\tilde{b}$ is expected to be quite light if the squared-mass of
$H_{\cal U}$ becomes negative. The VEV of  $H_{\cal U}$ will give a
positive contribution to the squared-mass of $\tilde{b}$, and hence
prevents $\tilde{b}$ from acquiring a nonzero VEV and breaking the
color gauge group. The low-energy theory in this case is a
two-Higgs-doublet model plus a charged color triplet scalar.
 

\begin{table}[t]
\centering \renewcommand{\arraystretch}{1.5}\small
\begin{tabular}{|c|c|c|c|c|}\hline
\parbox[t]{1.9cm}{Composite \\ \hspace*{.3cm} scalar} & 
\parbox[t]{2.1cm}{$\, $ \\ constituents  \\ $\, $ } & 
\parbox[t]{3.8cm}{$SU(3) \times SU(2) \times U(1)$\\ \hspace*{.5cm}
 representation} & 
\parbox[t]{2.8cm}{$\, $ \\ binding strength  \\ $\, $ } & 
\parbox[t]{2.7cm}{relative binding\\ for $\hg_1=\hg_2=\hg_3$} \\
\hline \hline\vbr\vbr 
$H_{\cal U}$ & $\overline{\cal Q}_+ {\cal U}_-$
& $({\bf 1,2,} +1/2)$ &  $\frac{4}{3} \hg_3^2 + \frac{1}{15}\hg_1^2$ &
1 \\ \hline\vbr\vbr $H_{\cal D}$ & $\overline{\cal Q}_+ {\cal D}_-$ &
$({\bf 1,2,} -1/2)$ &  $\frac{4}{3} \hg_3^2 - \frac{1}{30}\hg_1^2$ &
$0.93$ \\ \hline\vbr\vbr 
$\tilde{b}$ & $\ov{\cal Q}_+ {\cal Q}^c_-$ &
$({\bf 3,1,} -1/3)$ &  $\frac{2}{3} \hg_3^2 + \frac{3}{4}\hg_2^2 -
\frac{1}{60}\hg_1^2$ & $1 - \epsilon$ \\ \hline\hline\vbr\vbr
$\tilde{b}^\prime$ & $\ov{\cal U}_- {\cal D}^c_+$ & $({\bf 3,1,}
-1/3)$ &  $\frac{2}{3} \hg_3^2 + \frac{2}{15}\hg_1^2$ & $0.57$ \\
\hline\vbr\vbr 
$\tilde{b}^{\prime\prime}$ & $\ov{\cal Q}^c_- {\cal
L}_+$  & $({\bf {3},1,} -1/3)$ & $\frac{3}{4} \hg_2^2 +
\frac{1}{20}\hg_1^2$ & $0.57$ \\ \hline\vbr\vbr 
$\tilde{b}'''$ &
$\ov{\cal U}^c_+ {\cal E}_-$ &  $({\bf {3},1,} -1/3)$ &
$\frac{2}{5}\hg_1^2$ & $0.29$ \\ \hline $H_{\cal E}$ & $\overline{\cal
L}_+ {\cal E}_-$ & $({\bf 1,2,} -1/2)$ &  $\frac{3}{10}\hg_1^2$ &
$0.21$ \\ \hline\vbr\vbr 
$\tilde{q}''$ & $\overline{\cal L}_+ {\cal
D}_-$ & $({\bf 3,2,} +1/6)$ &  $\frac{1}{10}\hg_1^2$ & $0.07$ \\ \hline
\end{tabular} 
\parbox{15.cm}{\small \caption{\small Attractive scalar channels in
eight dimensions with chiral fermions.  We include an $\epsilon >0$ in
the $\tilde{b}$ channel to account for the  lifting of the degeneracy
due to the running coupling effect below $M_s$.
\label{8d_scalars} } }
\end{table}


\section{Four-fermion Operator Approximation}

\setcounter{equation}{0}

In the previous section we have studied the formation of bound states
using a most attractive channel approximation.  A more detailed study
of the bound state properties may be based on the following
considerations.

The higher-dimensional gauge interactions become strong at the
ultraviolet cutoff, and therefore the high-momentum gauge fields give
the dominant interaction between the fermions.  The picture described
in the previous section can be studied in a more quantitative manner
by approximating the  dynamics of the higher-dimensional gauge
interactions with an  effective theory involving four-fermion
operators suppressed by a scale  $\Lambda \sim M_s$\footnote{ If gauge
fields live in some additional dimensions where fermions do not
propagate, and those dimensions have sizes much smaller than $R$, then
one can first integrate out those additional dimensions and obtain the
four-fermion interactions suppressed by the scale of those
dimensions~\cite{Cheng:1999bg}. Even if these dimensions have size of
order $R$, the one gauge boson exchange is dominated by the ultraviolet and
leads to local, four fermion operators.  In the case where gauge
fields  and fermions propagate in the same dimensions, the
four-fermion  interactions generated by the gauge dynamics are
non-local.  Replacing them by local four-fermion operators is harder
to justify,  but analogous treatments in four dimensional gauge
theories  often work well empirically.}: 
\bear
\int d^D\! x \frac{-1}{2\Lambda^2} &\hspace*{-.4cm}  \left[\hg_3^2
\left( \overline{\cal Q}_+ \Gamma^\alpha T^r {\cal Q}_+
+\overline{\cal U}_- \Gamma^\alpha T^r {\cal U}_- \!\! +
\overline{\cal D}_- \Gamma^\alpha T^r {\cal D}_- \right)^2 + \hg_2^2
\bigg( \overline{\cal Q}_+ \Gamma^\alpha
\frac{\textstyle\vec{\sigma}}{\textstyle 2} {\cal Q}_+  +
\overline{\cal L}_+ \Gamma^\alpha  \frac{\textstyle
\vec{\sigma}}{\textstyle 2} {\cal L}_+ \bigg)^2 \right.  \nonumber \\
[2mm]
& \left. \hspace*{-2.2cm}  + \frac{\textstyle 3}{\textstyle 5}\hg_1^2
 \bigg(  \frac{\textstyle  1}{\textstyle  6}\overline{\cal Q}_+
 \Gamma^\alpha {\cal Q}_+
+ \frac{\textstyle  2}{\textstyle  3} \overline{\cal U}_-
\Gamma^\alpha {\cal U}_- - \frac{\textstyle  1}{\textstyle  3}
\overline{\cal D}_- \Gamma^\alpha {\cal D}_- - \frac{\textstyle
1}{\textstyle  2} \overline{\cal L}_+ \Gamma^\alpha {\cal L}_+ -
\overline{\cal E}_- \Gamma^\alpha {\cal E}_- \bigg)^2 \,\right]~,\!\!
\label{operators}
\eear where $\sigma$ are the Pauli matrices.

To be concrete, we study the $D=6$ case in this section.  The fermion
fields depend on the spacetime  coordinates $x^\alpha$, labeled by
$\alpha= 0,1,2,3, 5, 6$,  where $x^5$ and $x^6$ are compact, of size
$\pi R$.  The six-dimensional gamma matrices are given in terms of the
four-dimensional ones by, {\it e.g.},  
\be 
\Gamma^\mu = \left(
\begin{array}{cc} -\gamma^\mu       & 0 \\ 0          & \gamma^\mu
\end{array} \right) \; , \; \mu=0, 1, 2, 3, \;\;\; \Gamma^5= \left(
\begin{array}{cc} 0        & i{\bf I} \\ i{\bf I}       & 0
\end{array} \right) \;\; , \;\;\; \Gamma^6= \left( \begin{array}{cc} 0
& {\bf I} \\ -{\bf I}       & 0 \end{array} \right) \; ~, 
\ee 
and the 6-dimensional chiral projection operators are defined by 
\be
P_{\pm}\equiv\frac{1\pm \Gamma_7}{2} =\frac{1}{2}\left(
\begin{array}{cc} 1\mp \gamma_5        & 0 \\ 0       &  1\pm \gamma_5
\end{array} \right)  ~.  
\ee

The four-fermion operators (\ref{operators}) may be analyzed along the
lines presented in \cite{Cheng:1999bg}. The scalar channel operators
can be obtained after Fierz transformation,  
\be  
\int d^6x 
\frac{3}{2\Lambda^2} 
\hspace*{-.1cm}  \left[ c_{\cal U} \left(\overline{\cal Q}_+ {\cal
U}_-\right) \left(\overline{\cal U}_- {\cal Q}_+\right) + c_{\cal D}
\left(\overline{\cal Q}_+ {\cal D}_-\right) \left(\overline{\cal D}_-
{\cal Q}_+\right)  \right] + \cdots , 
\label{four}
\ee 
where $c_{\cal U}$, $c_{\cal D}$ are the binding strength for the
corresponding channels, which in the simplest approximation are
proportional to the value obtained in the MAC analysis,
($\frac{4}{3}\hg_3^2 + \frac{1}{15}\hg_1^2,\, \frac{4}{3}\hg_3^2 -
\frac{1}{30}\hg_1^2$,) and the ellipsis stand for vectorial and
tensorial four-fermion operators, which are irrelevant at
low energies, as well as four-fermion operators in the scalar channels
that do not produce light scalars.

The operators shown above give rise  to composite scalars whose
kinetic terms vanish at a scale $\sim M_s$. Therefore, these scalars
are physical degrees of freedom only below $M_s$.  We derive the
low-energy effective Lagrangian following the steps described in
\cite{Cheng:1999bg}. First,  the scalar self-energies and quartic
couplings are induced by the interactions with their
constituents. These may be  computed in the large-$N_c$ limit, where
only one fermion loop  contributes. Then the scalar fields may be
redefined to allow  canonical normalization of their kinetic terms.
This yields a six-dimensional effective action which  includes the
following terms involving scalars:  
\be  
- \int d^6 x \left[ V_6 +
\frac{\xi}{M_s} \left( \overline{\cal U}_- {\cal Q}_+ H_{\cal U} +
\overline{\cal D}_- {\cal Q}_+ H_{\cal D} + {\rm h.c.}  \right)
\right] ~, 
\ee 
where the effective potential is given by  
\be 
V_6 =
\frac{\lambda}{2M_s^2} \left( H_{\cal U}^\dagger H_{\cal U} + H_{\cal
D}^\dagger H_{\cal D}  \right)^2 + M_{H_{\cal U}}^2 H_{\cal U}^\dagger
H_{\cal U} + M_{H_{\cal D}}^2  H_{\cal D}^\dagger H_{\cal D} ~.  
\ee
The quartic and Yukawa couplings satisfy the usual 
NJL relation for large-$N_c$, 
\be 
\lambda = 2 \xi^2 ~.
\label{njl}
\ee 
The scalar squared-masses are strongly dependent on the cutoff,
but this does not affect the features important for the low-energy
theory,  namely their sign and relative sizes:  
\be 
\left(
M^2_{H_{\cal U}}, \ M^2_{H_{\cal D}} \right) \approx
\frac{16\pi^2F}{3N_c  }  \left( \frac{1}{c_{\cal U}}, \,
\frac{1}{c_{\cal D}} \right) - F^\prime\Lambda^2 ~, 
\ee 
where the
first term is the bare mass re-scaled by the wave function
renormalization and the second term comes from the fermion loop.  $F$
and $F^\prime$ are positive coefficients of order one that  may be
computed as in \cite{Cheng:1999bg}, by summing the loop integrals
corresponding to different KK modes.  The binding strength $c_{\cal
U},\, c_{\cal D}$ are proportional to the square of the
six-dimensional gauge couplings and have dimensions of  mass$^{-2}$
and are large in $M_s$ units, resulting in  $M^2_{H_{\cal U}} < 0$.

The minimum of $V_6$ is manifestly at $\langle H_{\cal U} \rangle \neq
0$ and $\langle H_{\cal D} \rangle = 0$.  Given that the
compactification scale is above the electroweak scale, the binding
strength needs to be adjusted close to the critical value where
$M^2_{H_{\cal U}}$ becomes negative. The binding strength  depends on
the strength of the higher-dimensional gauge couplings;  holding the
effective four-dimensional gauge couplings fixed, 
this can be adjusted by changing
the volume of the extra dimensions. The tuning that needs to be done
to keep the Higgs  light is not severe, since $M_s$ is less than a
factor of five higher than $1/R$ \cite{Cheng:1999fu}.

At scales below $1/R$ the two extra dimensions are integrated out, and
the four-dimensional effective theory is given by the Standard  Model,
(we describe the inclusion of three generations in  section 5,) with
the addition of a second Higgs doublet (the  $H_{\cal D}$ zero-mode).

In terms of the four-dimensional KK modes, the SM Higgs  $H_U \equiv
H_{\cal U}^{(0)}$ is a bound state of all the KK modes  of ${\cal
U}_-$ and ${\cal Q}_+$: 
\be 
H_U \sim \sum_{k=0}^{N_{\rm KK}-1}
\overline{\cal Q}_+^{(k)}  {\cal U}_-^{(k)} ~.  
\ee 
The coupling of
$H_U$ to each ${\cal Q}_+^{(k)}$ and  ${\cal U}_-^{(k)}$ mode is
suppressed by $\sqrt{N_{\rm KK}}$ compared with a four-dimensional top
condensate model.  Therefore, the top quark mass is also suppressed by
$\sqrt{N_{\rm KK}}$ compared with the $\sim 600$ GeV value expected in
the minimal four-dimensional top condensate model \cite{BHL} with a
TeV cutoff scale.

In the leading $N_c$ approximation, the  NJL relation (\ref{njl}) is
preserved after dimensional reduction. This implies  that the Higgs
boson mass, $M_h$, is also suppressed by $\sqrt{N_{\rm KK}}$ and is
given by $2m_t \approx 350$ GeV in the large $N_c$ limit.  This
suppression can also be understood as the volume factor of the compact
dimensions, ($N_{\rm KK} = V_{D-4} M_s^{D-4}$.)  Because the Higgs
doublet and the fermions live in extra dimensions, the
four-dimensional top Yukawa coupling and Higgs self-coupling are
related to the higher-dimensional ones by the volume factor: 
\be
\lambda_t = \frac{\xi}{\sqrt{V_{D-4} M_s^{D-4}}} \;\; , \; \;
\lambda_h = \frac{\lambda}{V_{D-4} M_s^{D-4}} ~.  
\ee 
By contrast,  in
top-quark seesaw models \cite{seesaw}, as well as in the model with
only $t_R$ in extra dimensions \cite{Cheng:1999bg}, the Higgs boson is
heavy, at the triviality bound, unless there is  large mixing among
scalars. 

The above discussion only includes the leading $N_c$ contribution,
{\it i.e.} fermion loops.  To get a more precise prediction of the top
and Higgs masses, one should also include the loop contributions from
gauge bosons and scalars. This can be done by computing the full
one-loop RG equations, and evolving the couplings from $M_s$ down to
the  electroweak scale.  The running of the quartic Higgs coupling
further decreases the physical Higgs boson mass. We study this effect
in the next section.

\section{Top and Higgs Mass Predictions}
\setcounter{equation}{0}

The more precise predictions of the top quark mass and Higgs mass
can be obtained from running the corresponding (four-dimensional)
couplings from the compositeness scale $M_s$, with the compositeness
boundary condition, $\yt,\, \hh \to \infty$ at $M_s$~\cite{BHL}, down
to low energies. The running is accelerated by the power-law between
the compositeness scale $M_s$ and the compactification scale
$M_c=1/R$, so the effect is significant even though the two scales are
not far apart. The low-energy predictions are governed by the infrared
fixed points of the RG equations \cite{Hill:1981sq}.  The infrared
fixed points are determined by the $\beta$-function coefficients
coming from the KK modes, which are different from those in the
four-dimensional Standard Model.

The one-loop RG equations for the (four-dimensional) SM gauge
couplings  above $M_c$ are given by 
\be 
16\pi^2 \frac{d g_i}{d \ln\mu} = 
N_{\rm KK}(\mu)\, b'_i g_i^3, 
\ee 
where $N_{\rm KK}(\mu)$ is
the number of KK modes below the scale $\mu$, [$ N_{\rm KK}(\mu) =
X_{\delta} (\mu R)^\delta, \, X_\delta =
\pi^{\delta/2}/\Gamma(1+\delta/2)$ in the continuous limit,] and
$b'_i$ are  
\bear 
b'_3 &=& -11 + \frac{2}{3}\, m\, n_g +
\frac{1}{2}\delta + \Delta_3 , \nonumber \\ [2mm] 
b'_2 &=&
-\frac{22}{3} +\frac{2}{3}\, m\, n_g+ \frac{1}{3}\delta +\frac{1}{6}
n_H + \Delta_2 ,\nonumber \\ [2mm] 
b'_1 &=& \frac{2}{3}\, m\, n_g +
\frac{1}{10} n_H +\Delta_1 , 
\eear 
$m$ is the number of fermion
components, ($m=4,\, 8$ for 6- and 8-dimensional chiral theories
respectively,) $n_g$ is the number of generations in the bulk (assumed
to be 1 throughout most of this section),  
$\delta=D-4$ is the number of extra
dimensions, $n_H$ is the number of light Higgs doublets, and
$\Delta_i,\, i=1, 2, 3$ represent the contributions from other
possible light composite scalars, ({\it e.g.}, a light $\tilde{b}$ in
eight dimensions contributes $1/6,\, 2/15$ to $\Delta_3$ and
$\Delta_1$ respectively.)

The one-loop RG equations for the top Yukawa coupling and the quartic
Higgs self-coupling are  
\bear 16 \pi^2 \frac{d\, \yt}{d \ln \mu} &=&
N_{\rm KK}(\mu) \, \yt\, \left\{ \frac{3(m+1)}{2} \, \yt^2
-\frac{24+4\delta}{3}\, g_3^2 -\frac{9}{4}\, g_2^2 -\frac{17}{20}\,
g_1^2 +\Delta_t \right\},  \\ [2mm] 
16 \pi^2 \frac{d\, \hh}{d \ln \mu}
&=& N_{\rm KK}(\mu) \, \Bigg\{ 12\hh^2 + 6 m \hh \yt^2 -6m \yt^4 -3\hh
\left(3g_2^2 +\frac{3}{5} g_1^2\right)  \nonumber \\ 
&&  +
\frac{3+\delta}{4}\left[ 2g_2^4 +\left(g_2^2+\frac{3}{5} g_1^2 \right)^2
\right] + \Delta_H \Bigg\}, 
\eear 
where $\Delta_t$ and $\Delta_H$
represent the contributions from other composite scalars.

\begin{figure}[ht]
\centerline{\epsfysize=9cm\epsfbox{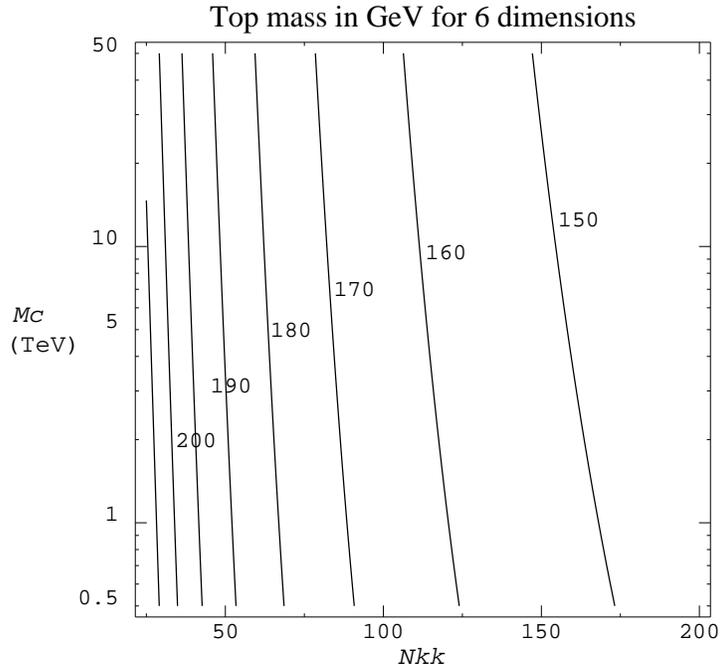}}
\begin{center}
\parbox{5.5in}{
\caption[] {\small The predicted top mass as a function of the number
of KK modes, $N_{\rm KK}$, and the compactification scale, $M_c$, in
the six-dimensional theory with $n_g=1$.
\label{mt6d}}}
\end{center}
\end{figure}

Combining the equations for $g_3$ and $\yt$, we obtain 
\be 
16\pi^2 \frac{d \ln (\yt/g_3)}{N_{\rm KK}(\mu) d \ln \mu} 
= g_3^2\, \left\{
\frac{3(m+1)}{2} \, \frac{\yt^2}{g_3^2}
-\left(\frac{24+4\delta}{3}+b'_3\right) -\frac{9}{4}\,
\frac{g_2^2}{g_3^2}  -\frac{17}{20}\, \frac{g_1^2}{g_3^2}
+\frac{\Delta_t}{g_3^2} \right\}.  
\ee 
If we neglect the contributions
from $g_2,\, g_1$, and $\Delta_t$, there is an infrared fixed point
for $\yt^2/g_3^2$ at $(48+8\delta+6b'_3)/(9m+9)$. For six dimensions,
assuming $n_g=1$ and $\Delta_3=0$, we have $\delta=2, \, m=4$, and
$b'_3=-22/3$. The infrared fixed point of $\yt/g_3$ is at  
\be
\left(\frac{\yt}{g_3}\right)_\ast = \frac{2}{3} \approx
\frac{0.8}{g_3(m_t)}.  
\ee 
$\yt/g_3$ decreases from $\infty$ at $M_s$
towards the fixed point in running down to low energies. How close
$\yt/g_3$ gets to the fixed point at $M_c$ depends on the ratio of
$M_s/M_c$, (or equivalently, the number of KK modes below $M_s,\, N_{\rm
KK}$.) Below the compactification scale $M_c$, the running follows the
four-dimensional SM RG equations.  The corresponding fixed point
becomes 
\be 
\left(\frac{\yt}{g_3}\right)_{{\rm SM}\ast} =
\sqrt{\frac{2}{9}} \approx \frac{0.6}{g_3(m_t)}, 
\ee 
so increasing
$M_c$ (while keeping $M_s/M_c$ fixed) will decrease the top mass
prediction, though the effect is small because of the slow logarithmic
running between $M_c$ and $m_t$. ($M_c$ should not be too large to
avoid extreme fine-tuning.) On the other hand, the $g_2$ and $g_1$
contributions will increase $\yt$ somewhat.  The value $0.8$ therefore
provides a rough lower bound on the  prediction of $\yt$ in this
case. The predicted top mass, $m_t=\yt v/\sqrt{2}, \, v=246$~GeV, for
a given $N_{\rm KK}$,  (or equivalently, $M_s/M_c$,) and
compactification scale $M_c$, can be obtained by numerically solving
the power-law and SM RG equations above and below $M_c$.  The result
is shown in Fig.~\ref{mt6d}.
The range of the parameters $M_c$ and $N_{\rm KK}$ should be such that
there is no excessive fine-tuning and there are enough KK modes to
produce non-perturbative strong dynamics, but not too many to cause SM
gauge couplings to reach the Landau pole.  In the figures we plot the
predicted masses for the range 0.5~TeV~$<M_c<$~50~TeV and $25<N_{\rm
KK}<200$. 

{}From Fig.~\ref{mt6d}, we see that the top quark mass predicted in
this theory is in agreement with the experimental value $174.3\pm
5.1$~GeV~\cite{PDG} with an uncertainty of $\sim 20\%$.

\begin{figure}[ht]
\centerline{\epsfysize=9cm\epsfbox{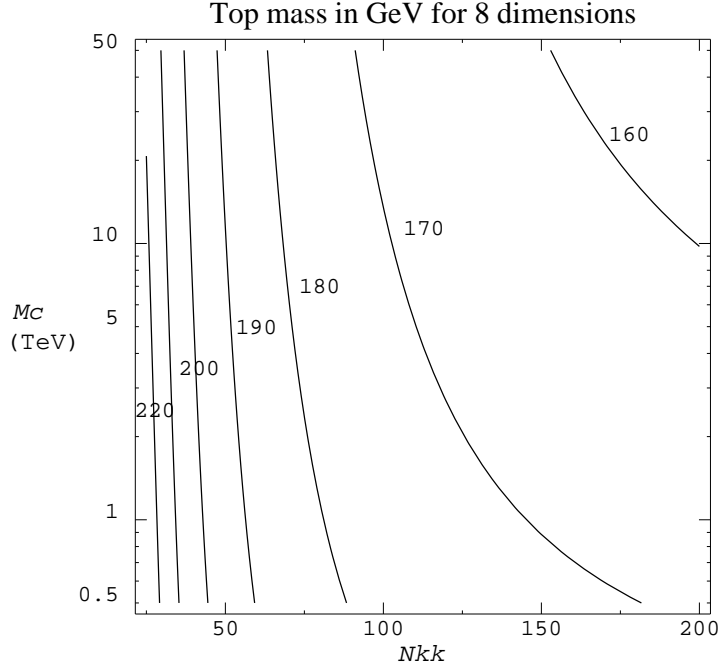}}
\begin{center}
\parbox{5.5in}{
\caption[] {\small The predicted top mass as a function of $N_{\rm
KK}$ and $M_c$ in the eight-dimensional theory with $n_g=1$.
\label{mt8d}}}
\end{center}
\end{figure}

In eight dimensions, the infrared fixed point for $\yt/g_3$ of the RG
equations between $M_c$ and $M_s$ (neglecting $g_2$, $g_1$ and $\Delta$'s)
is 
\be
\left(\frac{\yt}{g_3}\right)_\ast = \frac{\sqrt{58}}{9} \approx
\frac{1}{g_3(m_t)}, 
\ee 
so the predicted top mass is somewhat larger
compared with the six-dimensional case.  The numerical prediction is
shown in Fig.~\ref{mt8d}.
We can see that the prediction is also in good agreement with the
experimental value.

The Higgs mass is also controlled by the infrared fixed point
structure of the RG equations. Combining the RG equations for $\yt$
and $\hh$,  we obtain 
\bear 
16 \pi^2 \frac{d\ln( x_H)}{N_{\rm KK}(\mu) d \ln \mu} 
&=& \yt^2 \Bigg\{ 12x_H + 3(m-1)  -\frac{6m}{x_H}
+\frac{48+8\delta}{3}  \frac{g_3^2}{\yt^2}  -\frac{1}{\yt^2}
\left(\frac{9}{2}g_2^2 +\frac{1}{10}g_1^2\right)  \nonumber \\  
&&  +
\frac{3+\delta}{4 x_H \yt^4}\left[ 2g_2^4 +\left(g_2^2+\frac{3}{5}
g_1^2 \right)^2 \right] + \frac{\Delta_H}{ \hh^2} x_H
-\frac{2\Delta_t}{\yt^2}\Bigg\}, 
\eear 
where 
\be 
x_H \equiv
\frac{\hh}{\yt^2}.  
\ee 
If we neglect the contributions from the gauge
couplings and the $\Delta$'s, we find an infrared fixed point for
$x_H$ at 
\be 
12 x_H +3(m-1)-\frac{6m}{x_H}=0 \;\; \Rightarrow \;\;
x_{H\ast}=\frac{\sqrt{m^2+30 m+1}-m+1}{8}.  
\ee

\begin{figure}[t]
\centerline{\epsfysize=9cm\epsfbox{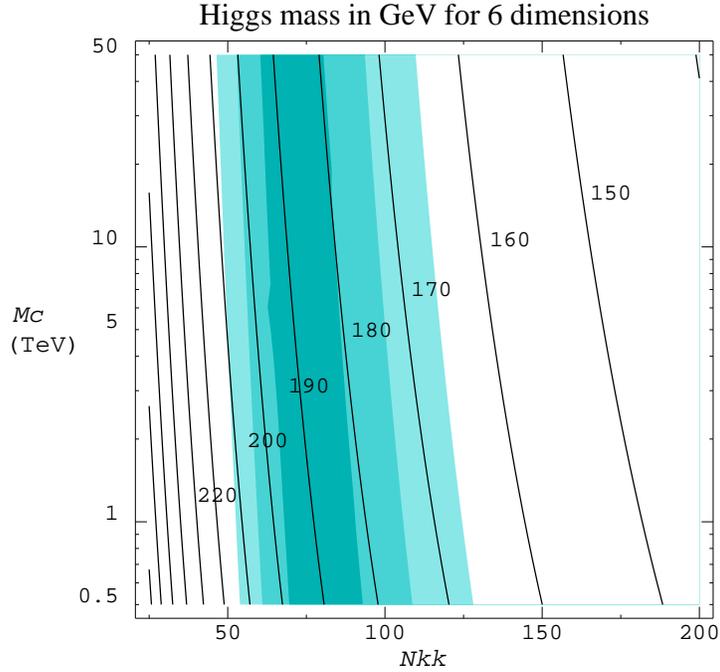}}
\begin{center}
\parbox{5.5in}{
\caption[] {\small The predicted Higgs mass as a function of $N_{\rm
KK}$ and $M_c$ in the six-dimensional theory with $n_g=1$.  The shaded
regions correspond to the top mass lying within 1--3 $\sigma$ (dark to
light) of the experimental value, $174.3\pm 5.1$~GeV.  
\label{mh6d}}}
\end{center}
\end{figure}
For six dimensions, $m=4$, $x_{H*}\approx 1.1$. The $(x_H-x_{H*})$
term is multiplied by a large coefficient in the RG equation,
therefore it approaches zero very rapidly. Numerically we find that
$\hh/\yt^2$ reaches $x_{H\ast}$ almost instantaneously below $M_s$. At 
lower energies, the $g_3^2/\yt^2$ term increases and it has a large
coefficient, so it is no longer a good approximation to neglect
it. This term reduces $x_H$ in running towards low energies. If we
assume that $g_3^2/\yt^2$  is constant and equal to its low-energy
value $g_3^2/\yt^2(m_t)$ for the correct top mass, the infrared fixed
point for $x_H$ becomes  
\be 
12 x'_{H\ast} +3(m-1)+\frac{64}{3}
\frac{g_3^2}{\yt^2}(m_t)- \frac{6m}{x'_{H\ast}}=0 \;\; \Rightarrow
\;\; x'_{H\ast} \approx 0.5 \; (\mbox{for } m=4).  
\ee 
Because $g_3^2/\yt^2$ is
smaller than $g_3^2/\yt^2(m_t)$ during the evolution, $x'_{H*}$
provides a rough lower bound on $x_H$ if we ignore the difference from
the SM running below $M_c$.  Therefore, for six dimensions we expect
\be 
0.5 \lae \frac{\hh}{\yt^2} \lae 1.1, \ee which translates to the
Higgs mass range \be 170\, {\rm GeV} \lae M_h=\sqrt{\hh} v \lae 260\,
{\rm GeV}.  
\ee

The dependence of the Higgs mass on $N_{\rm KK}$ and $M_c$ can also be
obtained numerically, and the result is shown for six dimensions  in
Fig.~\ref{mh6d}. Since the top mass has been determined
experimentally, we can obtain a better prediction of the Higgs mass
from the measured top mass.  In Fig.~\ref{mh6d}, we also show the
region of the parameter space which gives the top mass within
$3\sigma$ of the experimental value by the shaded area.  The
corresponding limit of the Higgs mass $M_h$ is 
\be 
165\, {\rm GeV} < M_h(\mbox{6-dim}) < 210\, {\rm GeV}.  
\ee
\begin{figure}[t]
\centerline{\epsfysize=9cm\epsfbox{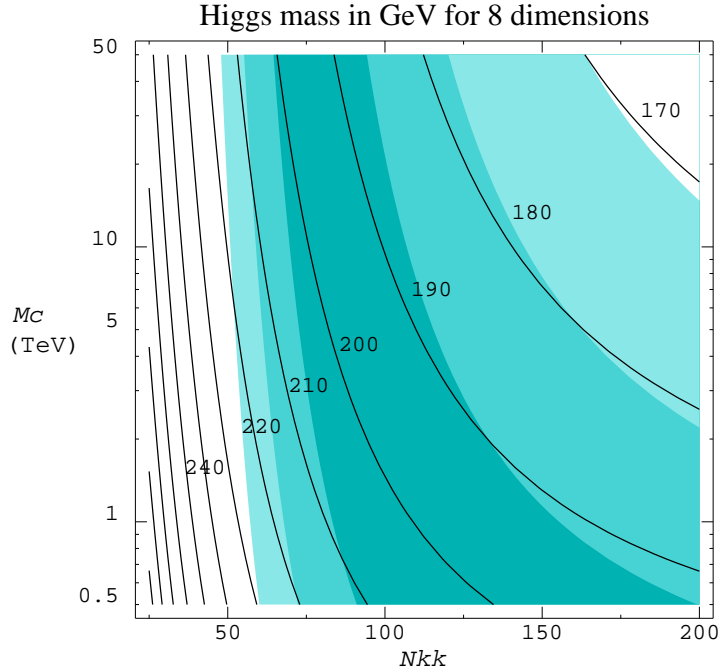}}
\begin{center}
\parbox{5.5in}{
\caption[] {\small The predicted Higgs mass as a function of $N_{\rm
KK}$ and $M_c$ in the eight-dimensional theory. The shaded regions
correspond to the top mass lying within 1--3 $\sigma$ (dark to light)
of the experimental value.  
\label{mh8d}}}
\end{center}
\end{figure}

Similar Higgs mass prediction can be obtained for the
eight-dimensional case. The fixed points $x_{H\ast}$ and $x'_{H\ast}$
are $1.3$ and $0.7$ in this case, which roughly correspond to 270 GeV
and 200 GeV  respectively. The numerical prediction for $M_h$ is shown
in  Fig.~\ref{mh8d}.

Due to the SM running below $M_c$, $M_h$ can in fact get below 200
GeV. The predicted Higgs mass in the eight-dimensional theory from
requiring a correct top mass within $3\sigma$ lies in the range 
\be
170\, {\rm GeV} < M_h(\mbox{8-dim}) < 230\, {\rm GeV}.  
\ee

As we emphasized in the introduction, the predictions of this section
have a  much more general validity than our particular mechanism for
triggering  electroweak breaking from Standard Model gauge dynamics in
extra dimensions.  They are a consequence of any theory where (1) the
field content is that  of the Standard Model, with the gauge bosons,
Higgs boson and one full  generation propagating in six or eight
dimensions, and (2) where the  higher-dimensional couplings
$\lambda_t,\lambda_h$ blow up in the ultraviolet, consistent with a 
composite Higgs boson. 

If the first two generations of fermions also propagate in extra
dimensions, there may be more light bulk bound states, which can
contribute to the power-law running  of the top Yukawa coupling and
the Higgs self-coupling.  As we will discuss in the next section, some
flavor breaking must be present so that only one Higgs gets a large
VEV. If we simply assume that there are no new bound states  even with
more generations propagating in the bulk,  the fixed points for
$\yt/g_3$ become  $1.15/g_3(m_t)\: (n_g=2)$, $1.3/g_3(m_t)\: (n_g=3)$,
for six dimensions, and  $1.3/g_3(m_t)\: (n_g=2)$,
$1.5/g_3(m_t)\: (n_g=3)$, for eight dimensions.  Contributions from
additional light scalars in the bulk can reduce the fixed points.
Consequently, more uncertainties are introduced in the top and Higgs
mass predictions, but we still expect the Higgs boson to remain rather
light.

\section{Flavor Symmetry Breaking}
\setcounter{equation}{0}

So far we have only discussed the case where the third generation of
fermions propagates in $D-4$ compact dimensions, without specifying
what happens with the other two generations. A possibility is that the
fermions of the first two generation are
four-dimensional~\cite{Carone:1999nz}, localized at some points in the
space of extra dimensions. In this case, there may be
(four-dimensional) bound states between the bulk fermions  of the
third generation and the four-dimensional fermions.  The binding force
of higher-dimensional scalars receives contributions from the extra
components of the gauge fields, and hence is stronger than the
four-dimensional ones at generic points in extra dimensions (away from
the orbifold fixed points) by $D/4$ in the lowest order approximation,
(as one can see from the Fierz transformations.)  The discussion in
the previous sections will hold if these four-dimensional bound states
are indeed heavy and do not appear in the low-energy theory.

A more natural option may be that all three generations fill the
$D$-dimensional spacetime, namely each of the $\CQ_+$, $\CU_-$,
$\CD_-$, $\CL_+$, $\CE_-$ fermions belongs to the fundamental
representation of a  global $U(3)$ symmetry. Therefore, the spacetime
configuration and  the Standard Model gauge interactions preserve a
$U(3)^5$ flavor symmetry.

As we showed in sections 2 and 3, the bound state with negative
squared-mass is the $\overline{\cal Q}_+ U_-$ scalar, which in the
case  of three generations belongs to the $(3,3)$ representation of
the  $U(3)_\CQ \times U(3)_\CU$ flavor symmetry. In other words, there
are  nine ``up-type'' Higgs doublets. In the absence of flavor
symmetry  breaking, these Higgs doublets are degenerate and obtain
VEV's that  break $U(3)_\CQ \times U(3)_\CU$ down to the diagonal
$U(3)$, leading to eight Nambu-Goldstone bosons in addition to the
ones eaten by the $W$ and $Z$. Clearly there is need for flavor
breaking, not only to give sufficiently large masses to these
Nambu-Goldstone bosons, but also to account for the various masses of
the  quarks and leptons. 

We now argue that any source of flavor breaking is likely to have a
large effect. Recall that the squared-mass of a composite Higgs
doublet is very sensitive to the strength of the  interaction between
its constituents. Therefore, some perturbative, flavor non-universal
interaction may easily tilt the vacuum in the direction where only one
Higgs doublet has a negative squared-mass.  This immediately
eliminates the unwanted Nambu-Goldstone bosons.

The flavor breaking can come from operators induced at the cutoff
scale $M_s$, such as the following four-fermion operators~\cite{ewsb},
\be 
\frac{\eta_{ij}}{M_s^{D-2}} \left(\overline{\cal Q}_+^i {\cal
U}_-^j\right) \left(\overline{\cal U}_-^j {\cal Q}_+^i \right) ~,
\label{non-universal}
\ee 
where $i=1,2,3$ labels the generations. If the attractive force is
enhanced in one channel (identified as the 3-3 channel) relative to
the others, then only one $H_{\cal U}$ (which couples to the third
generation) gets a VEV, while the squared-masses of other Higgs
doublets can stay positive. Note that given the sensitivity  of the
Higgs mass to the strength of the binding interaction, the other Higgs
doublets may be quite heavy even with a small splitting in the binding
strength. The flavor-changing effects induced by these scalars are
small if the scalar masses are large, or the $\eta_{ij}$ coefficients
approximately preserve some flavor symmetry~\cite{flavor}.

As in any theory with quantum gravity at the TeV scale,
flavor-changing effects become a problem if all possible
higher-dimensional operators consistent with the SM gauge symmetry are
induced with unsuppressed coefficients. One has to assume that the
problematic flavor-changing operators, such as $\Delta S =2$, are
suppressed by an underlying flavor symmetry or some other mechanism of
the fundamental short-distance theory.

With only one or two composite Higgs doublets in the low-energy
theory, the light quark and lepton masses can be generated by  certain
four-fermion operators induced at $M_s$.  To be specific, let us
discuss the $H_{\cal U}$ and  $H_{\cal D}$ bound states.  Note that
even though the squared-mass of $H_{\cal D}$ is likely to be positive
because the  $\overline{\cal Q}_+^3 {\cal D}_-^3$ channel is not
sufficiently strongly coupled, a  
\be 
\left(\overline{\cal Q}_+^3
{\cal U}_-^3 \right) \left(\overline{\cal Q}_+^3 {\cal D}_-^3 \right)
\label{quqd}
\ee 
operator would induce a VEV for $H_{\cal D}$. The important point
is  that operators such as 
\be 
\left(\overline{\cal Q}_+^3 {\cal
U}_-^3 \right) \left(\overline{\cal U}_-^i {\cal Q}_+^j\right)\; ,
\;\;   \left(\overline{\cal Q}_+^3 {\cal D}_-^3 \right)
\left(\overline{\cal D}_-^i {\cal Q}_+^j\right)\; , \;\;
\left(\overline{\cal Q}_+^3 {\cal D}_-^3 \right) \left(\overline{\cal
E}_+^i {\cal L}_-^j \right) 
\label{quuq}
\ee 
induce Yukawa couplings for the Higgs doublets \cite{MCHM}.  In
fact this choice of operators has a flavor structure that leads  in
the low-energy theory to a type-II two-Higgs doublet model, {\it
i.e.}, $H_{\cal U}$ gives masses to the up-type  quarks while $H_{\cal
D}$ gives masses to the leptons and down-type quarks.

Another possibility to prevent the first two generation forming light
bound states is that the fermions of different chirality are split in
the extra dimensions \cite{Arkani-Hamed:1999dc}.  Consider for example
the case that quarks and leptons propagate in $D=6$ dimensions, (four
infinite and two of radius $R$,) and there is one additional
transverse dimension with coordinate $x^7$ and  radius $R_T (>
M_s^{-1})$ smaller than $R$. Assuming that  the third generation is
localized at $x^7=0$, and the  other two generations are at $x^7 \neq
0$ with the $+$ and $-$ chiralities localized at different $x^7$, the
strength of the  attractive channels which involve the first two
generations is suppressed by the separation. In this case the spectrum
of bound states  is the same as the one described in section 2, namely
there is a single six-dimensional Higgs doublet, $H_{\cal U}$, with a
large Yukawa  coupling to the top quark, and a six-dimensional Higgs
doublet,  $H_{\cal D}$, with a large coupling to the bottom quark (and
$M_{H_{\cal D}}^2 > M_{H_{\cal U}}^2$.) The light fermion masses can
still arise from the operators (\ref{quqd}), (\ref{quuq}), with the
hierarchies explained by the distances between the fermions.

\section{A Comparison with Supersymmetry}
\setcounter{equation}{0}

Given the $SU(3)_C \times SU(2)_W \times U(1)_Y$ gauge structure of the
quark and lepton interactions, two crucial questions arise: why is the
gauge group broken spontaneously to $SU(3)_C \times U(1)_{EM}$, and why
does just one fermion, of charge 2/3, couple strongly to this symmetry
breaking.  Supersymmetric extensions of the Standard Model are known
to make significant progress on these questions, and in this section
we compare our mechanism with the case of supersymmetric electroweak
symmetry  breaking.

Our extra-dimensional approach shares certain features with
supersymmetric theories:  both extend spacetime symmetries and have
the breaking scale of this extra spacetime symmetry linked to the
scale of electroweak symmetry breaking. The gauge, quark and lepton
fields are extended to become representations of the larger spacetime
symmetry --- they propagate in superspace or in the extra-dimensional
bulk.  Furthermore, in both cases the dynamics which generates a
negative squared mass for the Higgs field is directly connected to 
the interaction which
leads to a heavy top quark. However, on closer inspection the
mechanisms are completely different and much insight is gained by
comparing the assumptions and accomplishments of these two  approaches.

Perhaps the largest difference is that in supersymmetric theories the
Higgs particles are added to the theory by hand, whereas in the
extra-dimensional theory they are automatically generated as quark
composites,  bound by the Standard Model gauge forces which become
strong in the bulk.  It is by no means obvious that Higgs doublets
need to be added in supersymmetric theories, since the scalar
superpartner of the lepton doublet has the right gauge quantum numbers
to be the Higgs boson. However, it has not proven possible to break
electroweak symmetry using only the sneutrino VEV --- one of the great
``missed opportunities'' of supersymmetry.

In supersymmetric theories it is very significant that the correct
pattern of electroweak symmetry breaking is triggered by the radiative
corrections induced by the large top quark Yukawa coupling. The theory
has many scalars: squarks, sleptons and Higgs bosons, yet only the
Higgs boson acquires a VEV. However, a large top quark Yukawa coupling
must be input into the theory by hand. Of course, experiment tells us
that the top quark is very heavy; but we  would like the theory to
explain why an up-type quark is heavy. It is  just as easy to
construct supersymmetric theories where the $\tau$ lepton  has a very
large Yukawa coupling rather than the top quark. In this case
supersymmetry predicts a different pattern of electroweak symmetry
breaking: $U(1)_Y$ is broken while $SU(2)$ survives as an unbroken
symmetry. Thus the success of supersymmetry is to correlate the
pattern  of electroweak symmetry breaking with the nature of the
heaviest fermion,  not to explain why a fermion is heavy. Contrast
this with the case that  the Standard Model gauge forces propagate in
6 or 8 dimensions. There  is no need to introduce an additional
non-gauge interaction by hand  for electroweak symmetry breaking. When
the gauge forces get strong, they bind a scalar Higgs  and
automatically induce a large Yukawa  coupling to an up-type quark.  No
interactions are needed beyond the  Standard Model gauge forces in the
extra dimensions -- it is as if the  gaugino interactions could
somehow induce electroweak symmetry breaking  and a large top quark
mass! Furthermore, there is a direct link between the gauge  quantum
numbers of a generation and the result that the very heavy  fermion is
an up type quark.

While supersymmetric radiative electroweak symmetry breaking employs a
heavy top quark effect, it does not predict the mass of the top
quark. In fact, a  very heavy top quark is not needed --- 50 GeV is
certainly sufficient. On  the other hand, the extra-dimensional
approach employs an NJL-like  mechanism. In four dimensions, this
would yield   a large top Yukawa coupling at the compositeness scale,
and unless this scale is very high (thereby necessitating an enormous
fine-tune), the top quark is much too heavy, $m_t \approx 600$ GeV.
However, the magic is that in extra dimensions,  the fundamental
higher-dimensional couplings   can naturally be large and yet be
consistent with the  more ``perturbative'' four-dimensional couplings
$g,\lambda_t,\lambda_h \sim 1$ due  to a moderate dilution factor from
the volume of the extra dimensions.  This is why our theories predict
naturally smaller top and Higgs masses.  In both types of theory there
is the possibility that the top quark mass  is determined by infrared
fixed point behavior of the renormalization  group equations for the
Yukawa coupling. In supersymmetry, quasi-fixed-point  behaviour leads
to a top quark mass $m_t \approx 205 \sin\beta$ GeV for $\tan\beta$
not too large~\cite{susyir}.  A correct top mass can be obtained for
$\tan\beta \sim 1.6$, which gives rise to a relatively light Higgs
boson. The lower bound on the Higgs mass from LEP II has ruled out
such a low $\tan\beta$ in the simplest Minimal Supersymmetric Standard
Model.  With extra dimensions, the need for criticality implies that
the top quark fixed point is relevant, even though it may  not be
reached, and leads to a correct prediction of the top quark mass,
although with considerable ${\cal O}$(20\%) uncertainties. This is a
very  significant result. A more precise prediction is frustrated by a
lack of control of the ultraviolet  behavior of the theory, implying
that one does  not know how closely the infrared fixed point is
approached. A correct prediction of the top quark mass in
supersymmetric theories requires additional  structure, such as
$SO(10)$ grand unification; for extra dimensions,  the correct
prediction is inherent to the mechanism of electroweak  symmetry
breaking induced by the Standard Model gauge interactions.

Both schemes share a common mystery: why is there a light Higgs boson?
In  the supersymmetric case, once the Higgs fields have been
introduced, it  is necessary to understand why they do not acquire a
gauge invariant mass  of order the Planck scale. In the case of extra
dimensions, the most  natural mass for the composite scalars is of
order the scale where the  gauge interactions get strong, 10 TeV for
example\footnote{ The lower bound on the compactification scale from
direct searches  of KK modes is below 500 GeV in the case of three
generations in the  bulk because the KK modes can be produced only in
pairs. Thus, the scale of compositeness could be in principle as low
as $\sim 1$ TeV.  However, indirect constraints from the electroweak
data are likely  to push this bound to the few TeV range.}.  For
supersymmetry, the best solution to this ``$\mu$ problem'' is to
introduce  a symmetry which forbids a bare Higgs mass in the
supersymmetric  limit, and arrange for the generation of the operator
$[\mu H_{\cal U}  H_{\cal D}]_F$ once supersymmetry is broken. For
extra dimensions, it is necessary to  assume that the strong gauge
dynamics is such as to bind the Higgs boson  close to criticality,
where its mass vanishes. We know of no symmetry  which can guarantee
this, so apparently a fine tune is necessary --- this  is clearly the
primary weakness of the extra-dimensional scheme. Perhaps  it is
accidental, or perhaps it results naturally from the  non-perturbative
gauge dynamics which we do not understand.

For both supersymmetry and extra dimensions, given the existence of a
light Higgs, the simplest schemes impose constraints on the mass of
the  Higgs boson. Unlike the Standard Model, the scalar quartic
coupling is  not a free parameter. In supersymmetric theories it is
related to the  electroweak gauge couplings in such a way that there
is a tree level  upper limit to the lightest Higgs mass of $M_Z$,
which gets increased by radiative  corrections to about 135 GeV.  With
dynamical electroweak symmetry  breaking one typically thinks of a
very heavy, or non-existent, Higgs  boson. However, the
extra-dimensional scheme has a light Higgs boson  because the
renormalization group equations of the dimensionally reduced  theory
has an infrared fixed point which is quickly reached, and which  sets
the self-coupling close to the square of the top Yukawa coupling.  The
expected range of the Higgs  mass in the simplest scenarios is in the
range 165 GeV to 230 GeV, and  has no overlap with the supersymmetric
case. In non-minimal theories  with extra light scalars, the
constraints on the Higgs mass are relaxed for both supersymmetry and
extra dimensions.

In supersymmetric theories one has the freedom to add Yukawa couplings
by  hand to describe the full mass spectrum and mixing matrices of the
quarks  and charged leptons. As in the Standard Model, it is easy to
construct a  realistic theory of flavor --- but at the expense of a
deeper  understanding, or any predictivity. In extra dimensions,
incorporating  flavor beyond the top quark mass is more challenging,
and potentially  more rewarding. For example, if all three generations
propagate in the  bulk there is a $U(3)^5$ flavor symmetry. The
composite Higgs multiplet  $H_{\cal U}$  transforms non-trivially as
(3,3) under $U(3)_Q \times U(3)_U$ and,  when it acquires a VEV, many
of its components become Goldstone bosons.   To avoid this it appears
that flavor, at least in part, may be a  phenomenon of the bulk.
Clearly, many geometrical  configurations are possible, but the
crucial ingredient must be  that flavor breaking is inextricably
linked to  spacetime symmetry breaking, which is not the situation
usually envisaged in  supersymmetric theories.

In both schemes, electroweak symmetry breaking is a manifestation of a
deeper spacetime symmetry breaking, so that the more fundamental
question  becomes the origin and nature of spacetime symmetry
breaking. In the case  of supersymmetry, the Standard Model is
protected to some degree from the  primordial supersymmetry breaking,
so that the question of mediating the  supersymmetry breaking to the
Standard Model becomes of paramount  importance to phenomenology. With
extra dimensions such protection is  absent --- the mediation of
spacetime symmetry breaking to the Standard Model occurs directly via
the KK spectrum of the excitations of the  Standard Model particles.  

In summary: extra dimensions offer a more predictive and constrained
mechanism for electroweak symmetry breaking than occurs in
supersymmetric  theories. The Standard Model gauge interactions create
a Higgs boson  as  a bound state of top quarks, induce it to acquire
a VEV, correctly  predict the top quark mass with ${\cal O}$(20\%)
uncertainties,  and predict a somewhat light Higgs boson in the
$165-230$~GeV range. It  is remarkable that the puzzle of electroweak
symmetry breaking may be  encoded in the Standard Model gauge
interactions and quantum numbers,  with no need for any extra
particles or interactions beyond those  required by extra-dimensional
propagation.  Given the very plausible assumptions we have made
regarding  the strong Standard Model gauge dynamics,  the only price
to be paid is a moderate tuning  to keep the composite Higgs boson
light.

\subsection*{Acknowledgements}

We would like to thank  C.T. Hill, K.T Matchev, M. Schmaltz, and
C.E.M. Wagner for  discussions. H.-C. Cheng thanks the Theory Group at
Lawrence Berkeley  National Laboratory for hospitality while the work
was initiated.  The work of N. Arkani-Hamed and L. Hall was supported
by DOE under contract DE-AC03-76SF00098 and by NSF under contract
PHY-95-14797.  H.-C. Cheng is supported by the Robert R. McCormick
Fellowship and by DOE Grant DE-FG02-90ER-40560. B.A. Dobrescu is
supported by DOE Grant DE-AC02-76CH03000.

\section*{Appendix: \ Vector-like Fermions}
\addcontentsline{toc}{section}{Appendix:  \ Vector-like Fermions}
\renewcommand{\theequation}{A.\arabic{equation}}
\setcounter{equation}{0}

In this Appendix we consider the case where the higher-dimensional
fermions are vector-like. This is always the case when  the number of
dimensions accessible to the fermions, $D$, is odd, but it also occurs
as a particular case for even $D$. 

Vector-like $D$-dimensional fermions may form all the bound states
discussed in section 2 as well as new ones.  In particular, the most
attractive channel is the gauge-singlet scalar made of $\ov{{\cal
Q}}{\cal Q}$. $H_{\cal U}=\ov{{\cal Q}}{\cal U}$ is still the most
attractive channel which transforms non-trivially under the SM gauge
group, but it is less strongly bound than the singlets $S_{\cal
Q}=\ov{{\cal Q}}{\cal Q}$ and $S_{\cal U}=\ov{{\cal U}}{\cal
U}$. Assuming $\hg_3=\hg_2=\hg_1$, the $S_{\cal Q}$ and $S_{\cal U}$
channels are stronger than $H_{\cal U}$ by $3/2$  and $8/7$
respectively, and hence will likely condense first.  The VEV's of
these singlets do not break any gauge symmetry.  However, they give
positive squared-mass to the Higgs, $H_{\cal U}$,  through their cross
interactions, (or equivalently, dynamical masses to the constituents
of the Higgs, $Q$, $U$.)  This may prevent the Higgs from acquiring a
VEV, jeopardizing the simple mechanism for electroweak symmetry
breaking.  It is a detailed question whether the Higgs can still
acquire a nonzero VEV in the presence of these singlets, and it is
hard to be estimated reliably with simple approximations.

One thing which can help electroweak symmetry breaking to occur is the
orbifold projection required to obtain the four-dimensional chiral
theory. Let us demonstrate it by an example with a simple
setup. Assuming that each higher-dimensional fermion has $2^{n+1}$
components, we can obtain a single four-dimensional chiral zero mode
by incorporating orbifold projections with $n$ ${\bf Z}_2$
symmetriess, with  the composite scalars $\ov{{\cal Q}}{\cal Q}$ and
$\ov{{\cal U}}{\cal U}$  being odd  under all $n$ ${\bf Z}_2$
symmetries. (By contrast, $H_{\cal U} =\ov{{\cal Q}}{\cal U}$ is even
under all ${\bf Z}_2$'s.) After decomposed into four-dimensional KK
modes, $S_{\cal Q}$ and $S_{\cal U}$  have no zero modes, and their
lowest modes will have a KK mass component of $\sqrt{n}/R$, which
makes their squared-masses less negative. In addition, their 
self-quartic-couplings will be enhanced by $(3/2)^n$, because their wave
functions are proportional to the Sine function in these $n$
directions and $\int_0^{2\pi R} dy (\sqrt{2} \sin y/R)^4 =3/2$.
Larger self-couplings and less negative squared-masses result in
smaller VEV's for $S_{\cal Q}$ and $S_{\cal U}$ and smaller
contributions to  the squared-mass of ${\cal H_U}$. Based on the
simplest one-loop effective potential estimate, one finds that for
$n>2$, the Higgs can still develop a nonzero VEV and break the
electroweak symmetry.

Although this analysis is hardly reliable and depends on how the extra
dimensions are compactified and the four-dimensional chiral fermions
are obtained, it shows that dynamical electroweak symmetry breaking is
not ruled out in this scenario. If electroweak symmetry breaking does
occur correctly, the low-energy theory will contain two Higgs
doublets, a color-triplet scalar  $\tilde{b}= \ov{{\cal Q}} {\cal
Q}^c$ discussed in section~2.2, and several gauge-singlet scalars,
$S_{\cal Q}=\ov{{\cal Q}}{\cal Q}$, $S_{\cal U} =\ov{{\cal U}}{\cal
U}$, and $S_{\cal D}=\ov{{\cal D}}{\cal D}$.


\vfil
\end{document}